\documentclass[referee,useAMS,usenatbib,usegraphicx]{mn2e}

\newcommand{\saxj}{SAX~J1808.4--3658}
\newcommand{\xtejb}{XTE~J1807--294}
\newcommand{\nudot}{\dot{\nu}}

\newcommand{\Dt}{\Delta t}

\newcommand{\pth}[1]{\left({#1}\right)}
\newcommand{\qpth}[1]{\left[{#1}\right]}

\newcommand{\Sin}[1]{\sin\pth{#1}}
\newcommand{\Cos}[1]{\cos\pth{#1}}

\newcommand{\Tstar}{T^\star}
\newcommand{\Porb}{P_{orb}}

\usepackage{subfigure}

\bibpunct{(}{)}{;}{a}{}{,}

\title[Precise orbital parameters of XTE~J1807--294]{Precise
  determination of orbital parameters in system with slowly drifting
  phases: application to the case of XTE~J1807--294}

\author[A. Riggio et al.]{ {A.
    Riggio,$^{1}$\thanks{E-mail:riggio@dsf.unica.it; disalvo@fisica.unipa.it;
  burderi@mporzio.astro.it}} T. Di   Salvo,$^{2}$ L.  Burderi,$^{1}$ 
  R. Iaria,$^{2}$ \newauthor 
  A. Papitto,$^{3,4}$ M.T. Menna,$^{4}$
  G. Lavagetto,$^{2}$ \\
  $^1$Dipartimento di Fisica, Universit\`a degli Studi di Cagliari,
  Cittadella Universitaria\\
  S.P. Monserrato - Sestu Km 0,700 09042, Monserrato (CA), Italy\\
  $^2$Dipartimento di Scienze Fisiche e Astronomiche, Universit\`a~di
  Palermo, Via Archirafi 36, Palermo I-90123, Italy\\
  $^3$Dipartimento di Fisica, Universit\'a degli Studi di Roma 'Tor
  Vergata', via della Ricerca Scientifica 1, 00133 Roma, Italy\\
  $^4$Osservatorio Astronomico di Roma, Sede di Monteporzio Catone,
  Via Frascati 33, Rome I-00040, Italy}

\begin{document} 
\date{}
\maketitle
\begin{abstract}
  We describe a timing technique that allows to obtain precise orbital
  parameters of an accreting millisecond pulsar in those cases in
  which intrinsic variations of the phase delays (caused e.g.\ by
  proper variation of the spin frequency) with characteristic
  timescale longer than the orbital period do not allow to fit the
  orbital parameters over a long observation (tens of days). We show
  under which conditions this method can be applied and show the
  results obtained applying this method to the 2003 outburst observed
  by RXTE of the accreting millisecond pulsar \xtejb~which shows in
  its phase delays a non-negligible erratic behavior.  We refined the
  orbital parameters of \xtejb~using all the 90 days in which the
  pulsation is strongly detected and the method applicable.  In this way we
  obtain the orbital parameters of the source with a precision more
  than one order of magnitude better than the previous available
  orbital solution, a precision obtained to date, on accreting
  millisecond pulsars, only for \saxj~analyzing several outbursts
  spanning over seven years and with a much better statistics.
\end{abstract}


\section{Introduction}

Low Mass X-ray Binaries (LMXB) are binary systems in which one of the
two stars is a neutron star (NS) with low magnetic field ($< 10^9
\mathrm{Gauss}$) which accretes matter from a low-mass ($< 1 \,
M_\odot$) companion star. According to the so-called recycling
scenario \citep[see for a review][]{Bhatta_91} millisecond radio
pulsars originate from LMXBs, where the accretion torques and the
relatively weak magnetic fields are able to spin up the NSs up to
millisecond periods. When the companion star stops transferring matter
to the NS, the NS can switch on as millisecond radio pulsar.

A striking confirmation of this scenario was the discovery in 1998 of
millisecond X-ray pulsars in transient LMXBs. The first LMXB observed
to show coherent pulsations at a frequency of $\sim 400$ Hz was the
well studied \saxj\ \citep{Wijnands_98, Chakra_98}.  Due to the weak
magnetic field of these sources, the chance to see a pulsed emission
from a LMXB is quite low. However, to date 8 LMXBs were discovered to
be accreting millisecond pulsars \citep{Wijnands_05}, and all of them
are in transient systems. They spend most of the time in a quiescent
state, with very low luminosities (of the order of $10^{31} - 10^{32}$
ergs/s) and rarely we go into an X-ray outburst with luminosities in
the range $10^{36} - 10^{37}$ ergs/s.  Indeed, of all these sources
only \saxj, which shows more or less regular outbursts every two
years, has been observed in outburst more than once with RXTE. All the
other sources have shown just one outburst in the RXTE era.

This fact makes the study of the timing properties of these sources
even more difficult, given that the duration of the observations is
not a matter of choice, but is conditioned by the duration of the
outbursts which, in turns, puts a constrain on the precision of the
parameters that we can derive. And this is also the reason why we have
to obtain all the information and the precision of the parameters we
need just using the available data.  In the case of accreting
millisecond pulsars, among the parameters of interest there are, of
course, the timing parameters, that are the orbital parameters and the
spin parameters. The orbital parameters can give us important
information on the binary system, its evolution, and even on the
nature (e.g. degenerate or not) of the companion star.  Also, a
precise orbital solution will be important for a precise determination
of the spin parameters, the spin period evolution and the accretion
torques acting onto the NS.

As already mentioned above, the knowledge with the maximum possible
precision of the orbital parameters is of fundamental importance in
itself and for a successive study of the spin and the spin variations.
The study of the frequency Doppler shift due to orbital motion of a
millisecond pulsar in a binary system is the first step to obtain an
estimate of the set of orbital parameters.  To refine this estimate
the next step is the study of the pulse phase shifts in order to
obtain differential corrections to the orbital parameters and
therefore a finer orbital solution. However, in some cases, not all
the data in which the coherent X-ray pulsations are visible can be easily
used to obtain the differential corrections. The pulse phase shifts
are frequently affected by intrinsic long-term variations and/or
fluctuations (probably caused by the accretion torques) which are
superimposed to the modulation due to the orbital motion of the
source, making the fit of the residual sinusoidal modulation much more 
complicated. Clear examples of these complex behaviors of the pulse phase
shifts in accreting millisecond pulsars can be found in
\citet{Burderi_06}, who analyze \saxj\ and find a big jump in the
pulse phase shifts of the fundamental harmonic of the pulse, and in
\citet{Papitto_07}, who analyze XTE~J1814--338 finding a modulation of
the pulse phase shifts, anti-correlated to the X-ray flux, superposed
on a general spin-down trend.

Of course, the presence of non negligible long-term variations of the
pulse phase shifts with time, make it very difficult to fit a long
dataset with differential correction as reported in
Eq.\ref{eq:phase_orb_doppler} in order to obtain a precise estimate of
the orbital parameters using all the available time-span. Using the
classical technique it is then necessary, in order to decouple the
orbital modulation from the proper fluctuations of the pulse phases,
to take into account %%%{\small temporarily eliminate} 
the latter in some way.  This is often impossible to obtain by fitting 
with a simple model, due to the observed complex behaviors and/or our poor
knowledge of the physics of the accretion torques.  In such cases we
are forced to fit differential corrections of the orbital parameters
on restricted time intervals, in which the proper
fluctuations or variations of the phase shifts can be safely
approximated with a simple model, e.g.\ a parabola. In these cases,
therefore, the precision of the orbital solution is limited by
the limited used timespan or by our ability to model intrinsic
phase variations. 

In this paper we describe a simple method which permits, under certain
conditions, to remove from the pulse phase shifts all the effects not
due to differential orbital parameters corrections.  We apply this
method to the source \xtejb, obtaining for the first time a complete
set of orbital parameters with a precision at least one order of
magnitude higher with respect to the previously available orbital
solution.

\section{Observations}

The millisecond X-ray pulsar \xtejb~was spotted by RXTE on February
21$^{st}$, 2003 \citep{Markward_atel_03a}. The source was observed
with PCA (Proportional Counter Array) and HEXTE (High Energy X-ray
Timing Experiment), the principal instruments on-board of RXTE
\citep{Jahoda_96}, from February 28 to June 22, 2003. \xtejb~was also
observed with other satellites such as XMM-Newton \citep{Campana_03,
Kirsch_04, Falanga_05}, Chandra \citep{Markwardt_iauc_03b} and
Integral \citep{Campana_03}. No optical or radio counterpart has been
reported.  \citet{Linares_05} have reported the presence of twin kHz
QPOs analyzing RXTE observation.

In literature several attempts have been done in order to derive the
orbital parameters of this source. \citet{Markward_atel_03a} give only 
source position and the orbital period. \citet{Kirsch_03}, analyzing an
XMM-Newton observation during the outburst, give an estimate of the
semi-mayor axis. The first complete set of orbital parameters was
reported by \citet{Campana_03}, and successively by \citet{Kirsch_04}
and by \citet{Falanga_05}, analyzing the same XMM-Newton observation. 
%%More recently \citet{Falanga_05}, analyzing the XMM-Newton 
%%observation but with a more simple approach give a new set
%%of orbital parameters. 
All these authors assumed as orbital period the period reported by 
\citet{Markward_atel_03a}.

Here we analyze all the archival RXTE observations of this source available
to date.  In particular, we use data from the PCA (proportional counter array)
instrument on board of the satellite RXTE. We use data collected in
GoodXenon packing mode, with maximum time and energy resolution
(respectively $1 \mu s$ and 256 energy channels). In order to improve
the signal to noise ratio we select photon events from PCUs top layer
and in the energy range 3-13 keV. Using the \textit{faxbary} tool
(DE-405 solar system ephemeris) we corrected the photon arrival times
for the motion of the earth-spacecraft system and reported them to
barycentric dynamical times at the Solar System barycenter. We use the
source position reported by \cite{Markwardt_iauc_03b} using the
\textit{Chandra} observation of the same outburst.

In order to test the goodness of the available orbital solution, we
correct the photon arrival times with the formula:
\begin{equation}
  \label{eq:time_orb_doppler}
  t_{em} \simeq t_{arr} - A \qpth{\Sin{m(t_{arr}) + \omega} + 
    \frac{\varepsilon}{2} \Sin{2m(t_{arr}) + \omega} - 
    \frac{3\varepsilon}{2}\Sin{\omega}},
\end{equation}
where $t_{em}$ is the photon emission time, $t_{arr}$ is the photon
arrival time, $A$ the projected semi-major axis in light seconds,
$m(t_{arr}) = 2\pi (t_{arr} - \Tstar)/\Porb$ is the mean anomaly,
$\Porb$ the orbital period, $\Tstar$ is the time of ascending node
passage, $\omega$ is the periastron angle and $\varepsilon$ the
eccentricity. We used the orbital parameters reported by
\cite{Kirsch_04}, adopting an eccentricity $\varepsilon = 0$ (see
Tab.~\ref{table1} for details).

We divided the whole observation in time intervals of 1/6 $\Porb$
length each and epoch-folded each of these data intervals with respect
to the spin period we reported in Tab.~\ref{table1}. The pulse phase
delays are obtained fitting each pulse profile with two sinusoidal
components, since higher-order harmonics were detectable in the folded
light curve.  We fixed the period of the sinusoids to 1 and 0.5 times
the spin period, respectively, and we used the phase of the
fundamental harmonic to infer the pulse phase shifts. In
Fig.~\ref{fig:fig1} we show the pulse phase delays obtained in this
way, where we have plotted only the pulse phase delays corresponding
to the folded light curves for which the statistical significance for
the presence of the X-ray pulsation was $> 3\sigma$.

In Fig.~\ref{fig:fig1} a residual orbital modulation it is clearly
visible, superimposed to an intrinsic long-term variation of the
phases, possibly similar to the erratic spin changes mentioned by  
\cite{Markwardt_04}.
%%On one hand in Fig.\ref{fig:fig1} an orbital modulation it is clearly
%%visible, on the other hand the erratic behavior is apparent,
%%\textbf{as already reported by \cite{Markwardt_04}}. 
It can be seen that this behavior has characteristic time scales of the 
order of several $\Porb$. In the next section we describe a method able 
to temporarily eliminate any long-term phase variations or fluctuations,
in order to easily fit the residual modulations of the phases at the
orbital period and to find a revised, more precise orbital solution.

\section{Differential corrections of the orbital parameters}

We propose here a simple method of analysis which allows to (temporarily) 
eliminate, or at least strongly reduce, the long-term variation and 
erratic behavior of the pulse phase shifts in order to derive a precise 
orbital solution.  
The residuals in the phase delays due to a non-perfectly corrected
orbital parameters is given by the expression:
\begin{eqnarray}
  \label{eq:phase_orb_doppler}
  \phi_{orb}(t) = P_{spin}^{-1} ( \quad & &\nonumber \\
  (\Sin{m(t) + \omega} + \frac{\varepsilon}{2} \Sin{2m(t) + \omega} - \frac{3\varepsilon}{2}\Sin{\omega} ) &dA \; &- \nonumber\\
  \frac{2\pi A}{\Porb} (\Cos{m(t) + \omega} + \varepsilon \Cos{2m(t) + \omega}) &d\Tstar &- \nonumber\\
  \frac{m(t) A}{\Porb} (\Cos{m(t) + \omega} + \varepsilon \Cos{2m(t) + \omega}) &d\Porb &+ \nonumber\\
  A (\frac{1}{2} \Sin{2m(t) + \omega} - \frac{3}{2}\Sin{\omega}) &d\varepsilon &+ \nonumber\\
  (A \Cos{m(t) + \omega} + \frac{\varepsilon}{2}
  \Cos{2m(t) + \omega} - \frac{3\varepsilon}{2}\Cos{\omega}) &d\omega &)
\end{eqnarray}
where $P_{spin}$ is the spin period with respect to the light curves
are folded and $dA$, $d\Tstar$, $d\Porb$, $d\varepsilon$, and $d\omega$
are the differential corrections to the orbital parameters (the
projected semi-major axis, the time of ascending node passage, the
orbital period, the eccentricity, and the periastron angle, respectively).

If we compute difference between the phase of two adjacent folded
light curves we obtain, for $\Delta\phi_{orb}(t_i)$, the expression:
\begin{eqnarray}
  \label{eq:orb_phase_diff}
  \Delta\phi_{orb}(t_i) = \phi_{orb}(t_{i+1}) - \phi_{orb}(t_i) = P_{spin}^{-1} (\quad & &\nonumber \\
  (\;2\Cos{m_i + \omega + m_\Delta/2}\Sin{m_\Delta/2} + \varepsilon \Cos{2m_i + \omega + m_\Delta} \Sin{m_\Delta}\;) & dA &+ \nonumber  \\
  \frac{4\pi A}{\Porb} (\;\Sin{m_i + \omega + m_\Delta/2}\Sin{m_\Delta/2} + \varepsilon \Sin{2m_i + \omega + m_\Delta} \Sin{m_\Delta}\;) & d\Tstar  &+ \nonumber \\
  (\;\frac{2 m_i A}{\Porb} (\Sin{m_i + \omega + m_\Delta/2}\Sin{m_\Delta/2} + \varepsilon \Sin{2m_i + \omega + m_\Delta} \Sin{m_\Delta}) - & & \nonumber \\
  \quad\frac{m_\Delta A}{\Porb} (\Cos{m_i + m_\Delta + \omega} + \varepsilon \Cos{2m_i + 2m_\Delta + \omega})\;) &d\Porb &+ \nonumber \\
  A \Cos{2m_i + \omega + m_\Delta}\Sin{m_\Delta} & d\varepsilon &- \nonumber \\
  (\;2\Sin{m_i + \omega + m_\Delta/2}\Sin{m_\Delta/2} +
  \varepsilon \Sin{2m_i + \omega + m_\Delta} \Sin{m_\Delta}\;) & d\omega &\;\;)
\end{eqnarray}
where we pose for simplicity $m(t_i) = m_i$ and $2\pi \Dt / \Porb =
m_{i+1} - m_i = m_\Delta$.

In this way, that is calculating the phase differences between two 
consecutive intervals instead of the phases, we apply a linear
filter to the pulse phase delays, for which we illustrate the
fundamental properties. We now use the term input to indicate the
original signal, that is the pulse phase delays vs.\ time, and output
to indicate the signal we obtain plotting the phase difference of each
interval with the following vs.\ time. When the input signal is a
sinusoid of period $P$, the output is another sinusoid with same
period but with different phase and amplitude. In Fig.~\ref{fig:fig2}
we report the gain $G$, that is the ratio of the amplitudes of the
output to the input signal, for a sinusoidal signal of period $P$.
The analytical expression for $G$ is: $G = 2\Sin{\pi\,\Dt/ P}$.  As
can be seen in Fig.~\ref{fig:fig2}, $G$ has the maximum for $P = 2\Dt$
coincident with the Nyquist frequency. For $P >> \Dt$ we have $G
\propto P^{-1}$. This filter is then a band-pass filter, limited at
high frequencies by the Nyquist frequency and at low frequencies we
can fix a limit at the period $P \simeq 12 \Dt$, at which the
amplitude is reduced to a half.  For period of $P \simeq 60 \Dt$ the
amplitude is reduced by a factor ten.

In particular for the case of \xtejb, instead of
plotting the obtained pulse phases as a function of time, we consider
the phase difference between each interval and the following one,
$\Delta\phi(t_i) = \phi(t_{i+1}) - \phi(t_i)$, in the hypothesis that
for each $i$ we have $t_{i+1} - t_i = \Dt$, where $\Dt$ is constant
during all the observation. In this way we obtain the phase shifts
shown in Fig.~\ref{fig:fig3} (the same of Fig.~\ref{fig:fig1} but plotting
the phase differences instead of the phases), where, as it is easy to see, 
the orbital modulation is still visible, but any long-term variation of 
the pulse phases is completely smoothed out. To produce this figure we
divided each pulse profile in 6 time bins (in other words we chose 
$\Delta t = \Porb / 6$) in order to maximize the signal to noise ratio. 
In fact, in this case the effect of the filter does not change the amplitude 
of the orbital modulation of the output with respect to the input
(from Fig.~\ref{fig:fig2} the Gain = 1 for $\Porb / \Delta t = 6$). 
%%, and the the errors on the phases do not depend on the chosen $\Delta t$.

Due to its linearity the application of this filter to a signal which
is the sum of several signals is equal to the sum of each filtered
signal. We can then analyze separately the response to the filter of
the Doppler shift due to the orbital motion without fear that the
erratic behavior of the source can alter the result. 
We note that, in cases like the one considered in this paper, where 
the orbital period is much shorter than long term variations of the 
phases (probably caused by accretion torques onto the neutron star), the 
phase variations induced by the orbital modulation can be studied 
independently of the phase variations induces by the spin evolution 
(see e.g. \citet{Burderi_07}, \cite{Papitto_07}). Therefore, our analysis does 
not introduce any error or approximation in the determination of the 
orbital solution.

\section{Results and Discussion} 

\begin{figure}
  \begin{center}
      \includegraphics[]{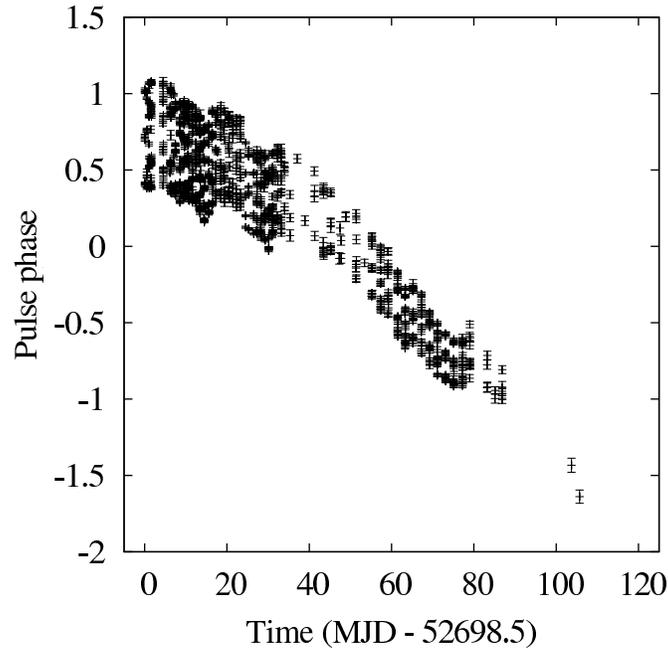}
      \caption{Plot of the pulse phase delays obtained by epoch folding the
        events barycentered with respect the orbital parameters reported by
        \citet{Kirsch_04} on time intervals of $\Porb/6$. It is
        clearly visible a residual orbital modulation superimposed to the
        long-term, sometimes erratic, behavior of the phases.}
      \label{fig:fig1}
  \end{center}
\end{figure}

\begin{figure}
  \begin{center}
      \includegraphics[]{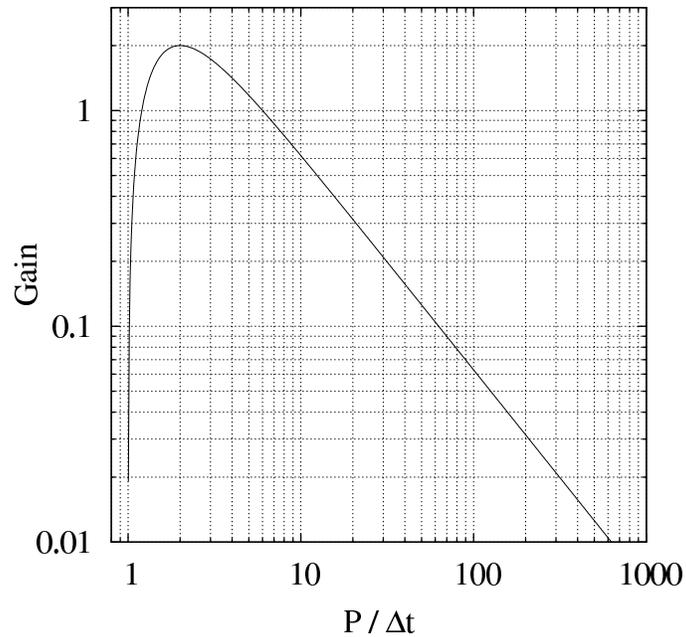}
      \caption{Plot of the gain of the filter, that is the ratio
        between the amplitude of a sinusoidal input signal of period
        $P$ and the amplitude of the output sinusoidal signal when the
        time distance between two adjacent points is $\Dt$.}
      \label{fig:fig2}
  \end{center}
\end{figure}

\begin{figure}
  \begin{center}
      \includegraphics[]{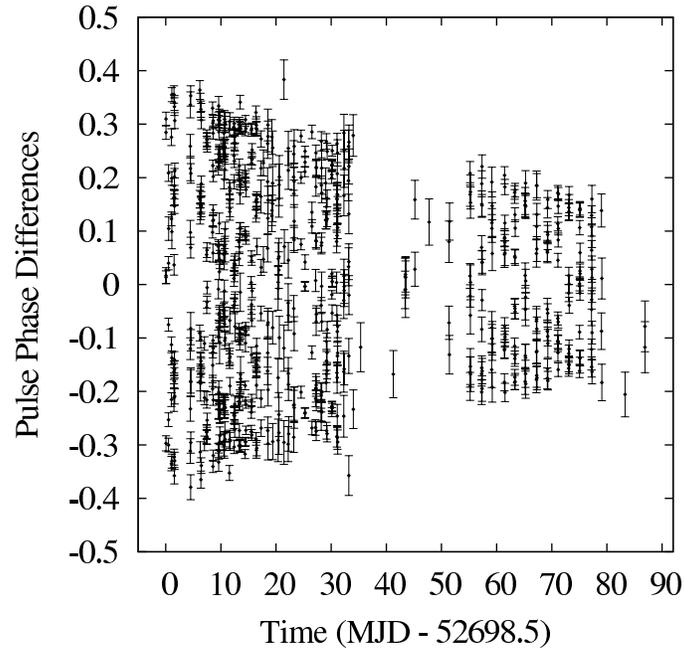}
      \caption{The same as Figure~\ref{fig:fig1} but plotting the pulse phase delays 
        differences (instead of the pulse phase delays, see text). As can be
        seen the erratic behavior and long-term variations result
        strongly reduced. The linear decrease of the amplitude is a
        clear sign of an error on the $\Porb$. }
        %% An estimation of this correction with the usual methods is, if not impossible,
        %%extremely difficult.}
      \label{fig:fig3}
  \end{center}
\end{figure}

\begin{figure}
  \begin{center}
      \includegraphics[]{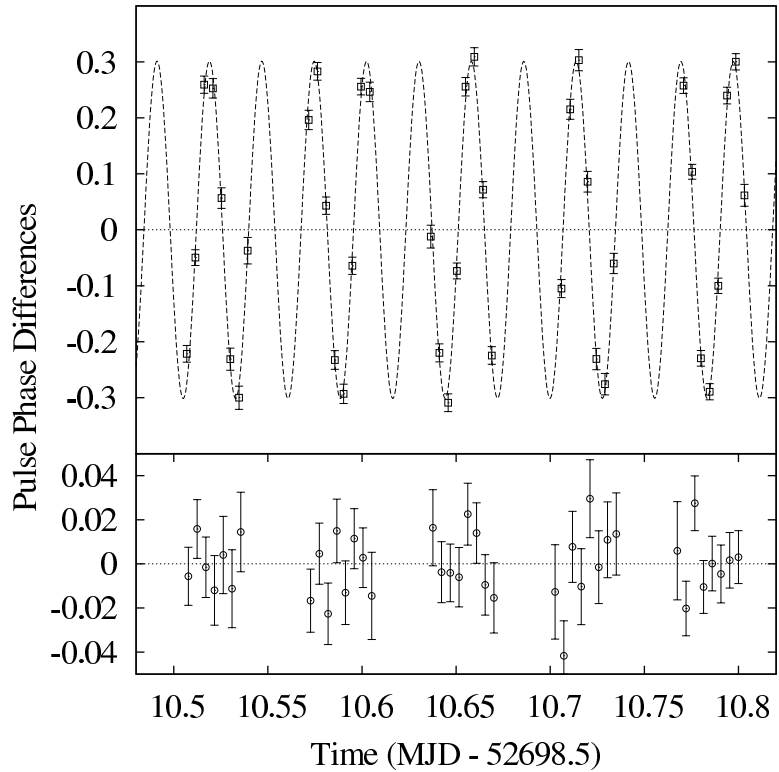}
      \caption{{\bf Top:} Detail of the pulse phase delays differences of
        Fig.\ref{fig:fig3} between the days 10 and 11 from the start
        of the observation. As can be seen there is only a sinusoidal
        modulation and there is no sign of the erratic behavior. The
        dotted line is the best-fit model described in
        Eq.\ref{eq:fit_simp_formula}. {\bf Bottom:} Post-fit residuals with
        respect to the best-fit sinusoidal model.}
      \label{fig:fig4}
  \end{center}
\end{figure}

We have applied the technique described above to the PCA data of
\xtejb. In particular, we have used the phase delays of
Fig.~\ref{fig:fig1} in order to calculate for each time interval the
phase difference with respect to the following interval, and these are
plotted vs.\ time in Fig.~\ref{fig:fig3}. We consider only phase
differences between contiguous time intervals and exclude the phase
differences between interval separated by gaps in time. The errors on
the phase differences are propagated summing in quadrature the errors
on the phases from which the difference is calculated, that is
$\sigma_{\Delta\phi(t_i)}^2 = \sigma_{\phi(t_{i+1})}^2 +
\sigma_{\phi(t_i)}^2$.  From the figure it is apparent that the long
term variation and the erratic behavior of the phase delays is now
completely smoothed out. We can therefore proceed to fit with
Eq.~\ref{eq:orb_phase_diff} these phase differences over the whole
period in which the coherent pulsation was detectable (about 90 days).  
In this way we
obtain a very good fit of the data. To show the goodness of the fit we
plot in Fig.~\ref{fig:fig4} (top panel) the phase differences between days 
10 and 11 from the start of the outburst; the dashed line is the best fit
sinusoidal modulation obtained from Eq.~\ref{eq:orb_phase_diff}. In the
bottom panel of the figure we show the post-fit residuals with respect to
the best fit sinusoidal modulation.

Eq.~\ref{eq:orb_phase_diff} is essentially a sum of sinusoidal terms
with period equal to $\Porb$ and $\Porb/2$. The latter are due only to
the eccentricity. Then, to test if the orbit shows an eccentricity we
epoch folded the light curves on a time interval $\Dt = \Porb/10$; this
reduces by about 40\% the Gain of the filter but gives the possibility 
to have a sufficient number of points to sample each period in order to
avoid aliasing phenomena. In fact, if we use, as before, intervals of
length $\Dt = \Porb/6$ this means that we sample the modulation at
$\Porb/2$ (eventually due to a non negligible eccentricity) with only
three points, and this can produce ambiguities in the results of the
fit. Using instead time intervals of length $\Dt = \Porb/10$, we
sample the modulation with period $\Porb$ with 10 points and the
modulation with period $\Porb/2$ with 5 points, which is, as we have
verified, a good compromise to get precise estimates of all the
orbital parameters.

To check that the long-term trend visible in Fig.~\ref{fig:fig1} has
been indeed eliminated by the technique described above we add to the
best-fit sinusoid a parabolic function to describe possible residuals
caused by the long-term phase variations. Hence we fit the phase 
differences with the expression:
\begin{equation}
  \label{eq:fit_formula}
  \Delta\phi(t) = a + b\,t + c\,t^2 + \Delta\phi_{orb}(t),
\end{equation}
where $a$, $b$ and $c$ are the coefficients of the
parabola. These coefficient can be expressed in terms of 
$\Delta\nu$, $\nudot$ and $\ddot{\nu}$, respectively, since 
$a \simeq -\Delta\nu \; 
\Delta t$, $b \simeq -\nudot/2 \; \Delta t$ and $c \simeq -
\ddot{\nu} / 3 \; \Delta t$. There is no evidence of residuals due
to the long-term behavior, and in the fit the $a$, $b$ and $c$
parameters result largely compatible with zero. This is due to
two factors: the first is that both $\Delta\nu$ and $\nudot$ are
attenuated by a factor $\Delta t\; (\simeq 5 \times 10^{-3})$ s, the
second is that the filter reduces the time dependence on these terms.
Moreover the orbit does not show an appreciable eccentricity,
$\varepsilon$, for which we find an upper limit (at 95\% c.l.) of $3.6
\times 10^{-3}$.  We also find that $d\Tstar$ and $d\omega$ result
perfectly correlated, as expected for a circular orbit.
  
\begin{figure}
  \begin{center}
    \includegraphics[]{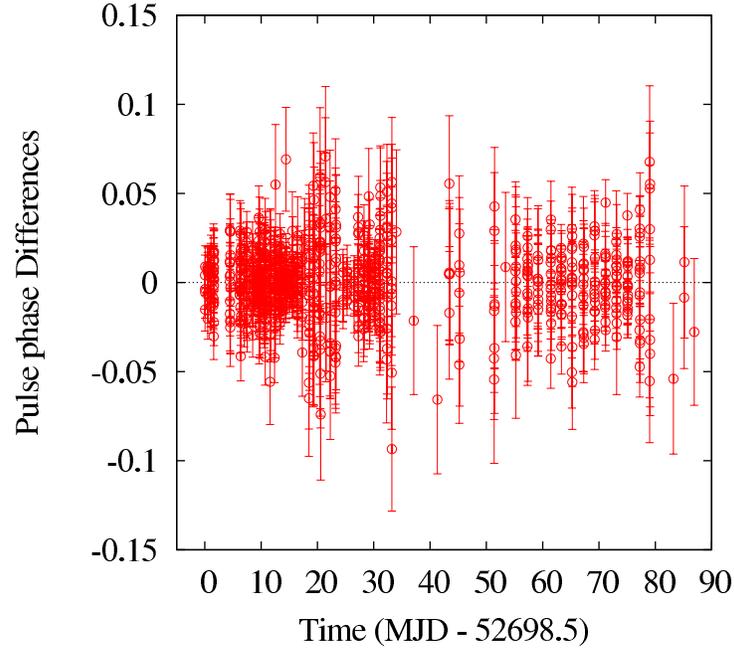}
    \caption{Plot of the pulse phase differences between contiguous intervals 
      using our orbital parameters. The sinusoidal modulation is no more visible
      and there is no sign of the erratic behavior and long-term
      variations. This implies that our technique perfectly smooths out 
      the erratic behavior and long-term variations.}
    \label{fig:fig5}
  \end{center}
\end{figure}

Due to these results we can safely make two assumptions: 1) the orbit
is circular; 2) we can safely describe the residuals simply with a
constant. We therefore epoch folded the light curves on a time interval 
$\Delta t = \Porb/6$ in order to have a better statistics, and fitted the
phase differences with the simpler formula:
\begin{equation}
  \label{eq:fit_simp_formula}
  \Delta\phi(t) = a + \Delta\phi_{orb}(t),
\end{equation}
where we fixed $d\varepsilon = d\omega = 0$.  We iterate this process
until no residual are observed. In this way we find a good fit,
corresponding to a $\chi^2 / \textit{d.o.f.}$ of $864.2 / 790$; 
the best fit parameters are reported in Tab.~\ref{table1}. 
In Fig.~\ref{fig:fig5}) we show the phase differences obtained correcting
the time series with our best-fit orbital solution. No orbital modulation
is visible in this plot, and the amplitude of the oscillation is now much
reduced with respect to that visible in Fig.~\ref{fig:fig3} corresponding
to the orbital solution given by \citet{Kirsch_04}.

To verify that our orbital solution is indeed better than the orbital
solution given by \citet{Kirsch_04} even during the times of the XMM
observation we performed the following check. 
We looked in the RXTE observations for a time interval close in time to 
the time of the XMM observation; unfortunately there is not a complete 
superposition between the RXTE and the XMM observation (that starts
at 52720.57 MJD and stops at 52720.68, for an exposure time of about
9.3 ks). We therefore took two RXTE observations ($80145-01-04-08$ and 
$80145-01-05-01$, the closest continuous observations to the XMM observation,
which cover the time interval from 52718.94 MJD to 52719.11 MJD), corrected 
them alternatively with the Kirsch solution and our solution, respectively, 
and then performed a folding search around the expected value of the spin
period. While the Kirsch solution gives a  peak in the $\chi^2$ curve of 
about $150 - 200$, our solution gives a peak in the $\chi^2$ curve 
of about 1000, demonstrating that the periodic signal revealed on the
time series corrected with our orbital solution is much stronger even
in a time interval as close as possible to the XMM observation.

\begin{figure}
  \begin{center}
    \includegraphics[]{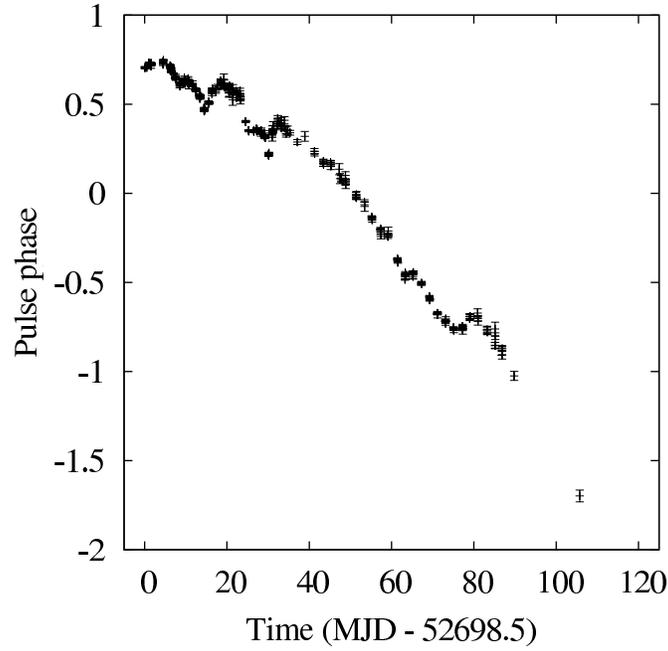}
    \caption{Pulse Phase delays as a function of time of \xtejb~obtained 
      correcting the time series with our best fit orbital solution. The light 
      curves are folded on a time interval of length $\Porb$. It is now apparent
      again the erratic behavior of the phases on characteristic timescales of
      at least several dozen of $\Porb$.}
    \label{fig:fig6}
  \end{center}
\end{figure}

The method described above that we used to determine a precise orbital
solution for \xtejb\ does not allow any study of the spin frequency and
its derivative, since the long-term phase variations, which indeed give 
information on possible variations of the spin frequency during the 
outburst, are eliminated when we calculate the phase differences.
Therefore, in order to perform a timing study of the spin frequency,
we have to correct the entire time series of the RXTE observations
with our best-fit orbital solution, and ri-calculate the phase delays, 
which are shown in Fig.~\ref{fig:fig6}. 
As it can be seen from the figure, the strong sinusoidal modulation visible
in Fig.~\ref{fig:fig1} is no more present. Moreover, our more precise orbital
solution allows us to clearly detect the coherent pulsations up to 104
days since the start of the observation (June 12) with a detection
confidence level of $3.6\; \sigma$, while between May 26 and June 10,
although the source is still detectable, the pulsations are no more
significantly detected. Long-term variations (a parabolic trend on which 
erratic fluctuations are superimposed) of the phases are again visible in 
this figure and are probably determined by the presence of a spin frequency
derivative and/or an error in the spin frequency used to obtain the folded 
pulse profile, as well as by fluctuations of the phases on shorter 
timescales possibly related to phase shifts in the neutron star surface caused
by variations of the X-ray flux. The discussion of these effects is beyond 
the scope of this paper and will be presented elsewhere \citep{Riggio_07}.

\section{Conclusions}

We have described a simple technique that permits to drastically
reduce the presence of erratic behavior and long-term intrinsic
variations of the pulse phase delays of the source, thus allowing to
fit the residual orbital modulation of these phase delays, caused by errors
in the previously reported orbital parameters, on a very long time-span and
to obtain a much more precise measure of the orbital parameters. We applied this
technique to the source \xtejb, which shows the longest X-ray outburst
observed by RXTE from an accreting millisecond pulsar. In this way we
can fit the residual modulation of the phase differences over the
whole time-span in which the coherent pulsations were significantly
detected (about 100 days from the start of the outburst), obtaining a
set of orbital parameters with a precision that is at least one order
of magnitude better than the previously published orbital solutions
for this source.

Once a good orbital parameters set is known, a detailed discussion of the 
spin period and its derivative is possible. However, this source also shows 
erratic fluctuations of the phases that are anticorrelated to variations 
in the X-ray flux, in a way that is similar to what is found by \cite{Papitto_07} 
for the source XTE~J1814--338. A detailed discussion of these effects and 
a determination of the spin frequency and its derivative in \xtejb\ will
be presented elsewhere \citep{Riggio_07}.

\thanks{We acknowledge the use of RXTE data from the HEASARC public archive.
  This work was supported by the Ministero della Universit\`a e della Ricerca (MiUR)
  national program PRIN2005 2005024090$\_$004.}

% Stile bibliografico e bibliografia
\bibliography{ms}

\setcounter{table}{0}
\begin{table*}
  \begin{minipage}{110mm}
    \caption{Orbital Parameters for XTE J1807-294.}
    \label{table1}
    \begin{tabular}{@{}lcc}
      \hline
      Parameter & Other Works & This Work \\
      \hline
      Orbital period, $\Porb$ (s) & 2404.45(3)$^a$ & 2404.41665(40) \\
      Projected semi-major axis, $a_x \sin i$ (lt-ms) & 4.8(1)$^b$ & 4.819(4) \\
      Ascending node passage, T$^\star$~$^a$ (MJD) & 52720.67415(16)$^b$ & 52720.675603(6) \\
      Eccentricity, e  & - & $<$ 0.0036 \\
      Spin frequency, $\nu_0$ (Hz) & 190.623508(15)$^b$ & 190.62350694(8)$^c$ \\
      \hline
    \end{tabular}   
    \medskip
    Errors are intended to be at $1\sigma$ c.l., upper limits are given at 95\% c.l.\\
    $^a$\cite{Markwardt_iauc_03b}.\\
    $^b$\cite{Kirsch_04}.\\
    $^c$\cite{Riggio_07}.\\
  \end{minipage}
\end{table*} 

\end{document}